\documentclass[aps,prl,twocolumn,groupedaddress,showkeys]{revtex4}
\usepackage{graphicx}
\usepackage{amsmath}
\pdfoutput=1

\begin{document}

\title{Mapping dynamical heterogeneity in structural glasses to
  correlated fluctuations of the time variables}

\author{Karina E. Avila} 
\author{Horacio E. Castillo}
\email[]{castillh@ohio.edu}
\affiliation{Department of Physics and Astronomy, Ohio University, Athens, OH, 45701, USA}

\author{Azita Parsaeian}
\affiliation{Materials Research Center, Northwestern University, Evanston, IL 60208-3108, USA\\}%

\date{\today}

\begin{abstract}
{\em Dynamical heterogeneities\/} -- strong fluctuations near the
glass transition -- are believed to be crucial to explain much of the
glass transition phenomenology. One possible hypothesis for their
origin is that they emerge from soft (Goldstone) modes associated with
a broken continuous symmetry under time reparametrizations. To test
this hypothesis, we use numerical simulation data from four
glass-forming models to construct coarse grained observables that
probe the dynamical heterogeneity, and decompose the fluctuations of
these observables into two {\em transverse\/} components associated
with the postulated time-fluctuation soft modes and a {\em
  longitudinal\/} component unrelated to them. We find that as
temperature is lowered and timescales are increased, the time
reparametrization fluctuations become increasingly dominant, and that
their correlation volumes grow together with the correlation volumes
of the dynamical heterogeneities, while the correlation volumes for
longitudinal fluctuations remain small.
\end{abstract}

%
%
%

\preprint{NSF-KITP-10-095}


\maketitle


For systems in the vicinity of the glass transition, experiments and
simulations have shown the emergence of spatially heterogeneous
dynamics (SHD): mesoscopic regions relax either much faster or much
slower than neighboring regions~\cite{Debenedetti2001, Ediger2000,
  Kob1997, Weeks2000, Russell2000, Toninelli2005, Cipelletti2003,
  Keys2007}. SHD is believed to be crucial to the understanding of
non-exponential relaxation, the breakdown of the coupling between
translational diffusion and viscosity, and even possibly the slowdown
of the dynamics itself~\cite{Debenedetti2001, Ediger2000}. The origin
of SHD is still uncertain, in part because of the lack of direct
microscopic tests to attempt to disprove proposed
theories~\cite{Garrahan2002, Lubchenko2007, Toninelli2005,
  Castillo2003}. Here we apply one such test~\cite{Jaubert2007} for
the hypothesis that SHD is associated with fluctuations in the time
variable~\cite{Castillo2003, Chamon2002, Castillo2002}, and find that
our molecular dynamics data are consistent with the hypothesis. This
test can also be applied to particle tracking experimental data in
colloidal~\cite{Weeks2000} and granular systems~\cite{Keys2007}, thus
allowing to investigate a possible unified explanation of SHD in
diverse systems. Our results highlight that non-trivial correlation
functions in the time domain contain useful information for the
understanding of SHD.

As a glass-forming liquid approaches the glass transition, its
relaxation time and viscosity grow by many orders of magnitude, until
the system can no longer equilibrate in laboratory timescales, i.e. it
has entered the glass state~\cite{Debenedetti2001}. In equilibrium,
the correlation function $C(t,t_w)$ between the states of the system
at the {\em waiting time\/} $t_w$ and the {\em final time\/} $t$
depends only on $t-t_w$, but if the system is out of equilibrium, it
may display {\em aging\/}, i.e. a nontrivial dependence on both $t$
and $t_w$. Dynamical heterogeneity can be probed by defining a coarse
grained {\em local\/} two-time correlation $C_{\bf r}(t, t_w)$, which
probes how much each
individual region of the sample has changed between time $t_w$ and
time $t$. ``Fast regions'' have small values of $C_{\bf r}(t, t_w)$
and ``slow regions'' have values of $C_{\bf r}(t, t_w)$ closer to
1. Thus {\em the fluctuations of $C_{\bf r}(t, t_w)$ represent the
dynamical heterogeneity, and theories attempting to explain SHD should
be able to explain those fluctuations.} One of the proposed mechanisms
for the origin of dynamical heterogeneity postulates that they are
associated with {\em local fluctuations in the time
  variable\/}~\cite{Castillo2003, Chamon2002, Castillo2002,
  Castillo2007, Castillo2008, Parsaeian2008a, Parsaeian2008,
  Parsaeian2009}, $t \to h_{\bf r}(t)$, i.e.
\begin{equation}
  C_{\bf r}(t, t_w) = C(h_{\bf r}(t), h_{\bf r}(t_w)), 
\label{eq:C_goldstone}
\end{equation}
where $C(t, t_w) \equiv C_{\mbox{\scriptsize global}}(t,t_w)$ is the {\em
  global\/} two-time correlation. This proposal originated in analytic
calculations in spin glass models in the long time limit that showed
the presence of a broken continuous symmetry under reparametrizations
of the time $t \to h(t)$~\cite{Chamon2002, Castillo2008}, which should
give rise to the presence of Goldstone modes as described by
Eq.~\ref{eq:C_goldstone}.  Indirect evidence in favor of the presence
of this kind of fluctuation in atomistic models of glasses has been
presented in~\cite{Castillo2007, Parsaeian2008a, Parsaeian2008,
  Parsaeian2009}. In the present work, we introduce a more direct
test, based on decomposing fluctuations into a {\em transverse\/} part
satisfying Eq.~(\ref{eq:C_goldstone}) and a {\em longitudinal\/} part
containing all other fluctuations~\cite{Kardar2007}. This procedure
allows one to separately quantify the strength and spatial
correlations of both kinds of fluctuations, as a function of
temperature and timescales, for a variety of glass-forming models, and
is easily applicable to experimental data in glassy colloidal and
granular systems.


To probe fluctuations in structural glasses, we use~\cite{Castillo2007}
$C_{\bf r}(t, t_w) = \frac{1}{N(B_{\bf r})} \sum_{{\bf r}_j(t_w) \in
  B_{\bf r}} \cos({\bf q} \cdot ({\bf r}_j(t) - {\bf r}_j(t_w)))$. Here
${\bf r}_j(t)$ is the position of particle $j$ at time $t$, $B_{\bf r}$
denotes a small coarse graining box around the point ${\bf r}$, and
the sum runs over the $N(B_{\bf r})$ particles present in the coarse
graining box at the waiting time $t_w$. The {\em global\/} correlation
function $C(t,t_w)$, defined by
extending the average to all of the $N$ particles in the system, is
the self part of the intermediate scattering function. We have chosen
the wavevector ${\bf q}$ to be at the main peak of the static
structure factor $S({\bf q})$ for each system. We performed classical
Molecular Dynamics simulations of systems of $N$ particles ($1000 \le
N \le 8000$) that were equilibrated at high temperature $T_i \gg T_g$,
then instantaneously quenched to a final temperature $T$ and allowed
to evolve for times several orders of magnitude longer than their
typical vibrational times~\cite{Castillo2007, Parsaeian2008a,
  Parsaeian2008, Parsaeian2009}.  We generated eight datasets by
simulating four atomistic glass-forming models~\cite{Berthier2009}: an 80:20 mixture of
particles interacting via Lennard-Jones (LJ)
potentials~\cite{Kob-Barrat-aging-prl97, Castillo2007} (dataset C), an
80:20 mixture of particles interacting via purely repulsive
Weeks-Chandler-Andersen (WCA) potentials~\cite{Parsaeian2009}
(datasets D-H), and short (10-monomer) polymer
systems~\cite{Parsaeian2008a} interacting via either LJ potentials
(dataset A) or via WCA potentials (dataset B). Nearest neighbors along
the polymer chains are held together by FENE anharmonic spring
potentials~\cite{Parsaeian2008a}.  The ratio of the final temperature
$T$ to the Mode Coupling critical temperature
$T_c$~\cite{Bengtzelius1984} was $T/T_c \sim 0.9$ for datasets A-D,
$T/T_c = 1.10$ for datasets E-F and $T/T_c = 1.52$ for datasets
G-H. For datasets F and H, the samples were in equilibrium, but for
all the others the samples were aging. Each dataset includes between
100 and 9000 independent runs with the same parameters.


\begin{figure*}[ht]
  \begin{center}
    \includegraphics[trim=0.7in 7.5in 0in 0.75in, clip=true]{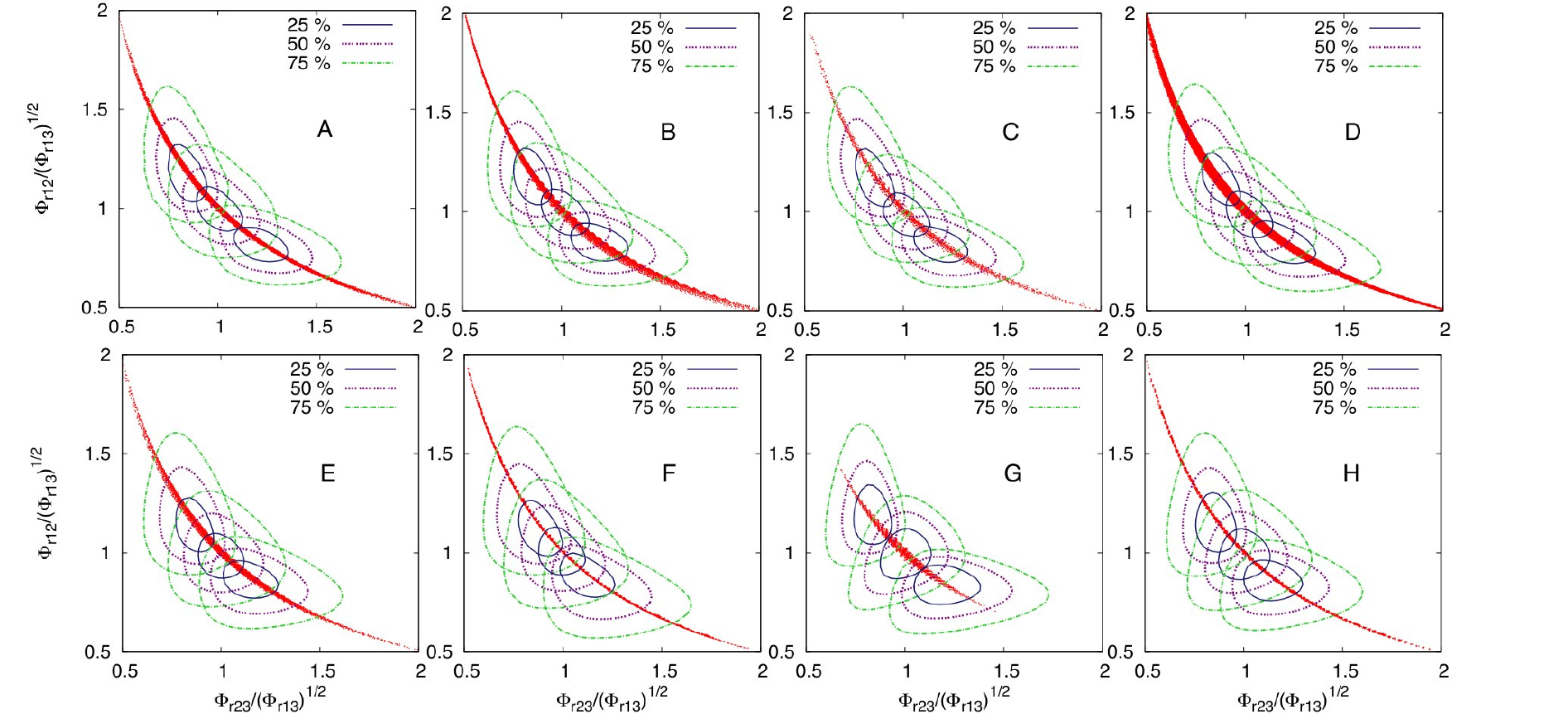}
  \end{center}
  \caption{(Color online) 2D contours of constant joint probability
    density $\rho(X_{\bf r}, Y_{\bf r}) = \rho(\Phi_{23, {\bf
        r}}/(\Phi_{13, {\bf r}})^{1/2}, \Phi_{12, {\bf r}}/(\Phi_{13,
      {\bf r}})^{1/2})$, computed using coarse graining boxes containing
    125 particles on average. Each set of three concentric contours is
    chosen so that they enclose $25\%, 50\%$ and $75\%$ of the total
    probability. Each panel from A to H contains results from the
    corresponding dataset, for $(X, Y) \approx (0.80, 1.25), (1.00,
    1.00)$ and $(1.25, 0.80)$, with the times chosen as late as possible
    within each dataset. The global values $(X(t_1, t_2, t_3), Y(t_1,
    t_2, t_3))$, for {\em all\/} times $t_1 > t_2 > t_3$ in each dataset,
    are shown with red points.}
    \label{fig:all}
\end{figure*}

\begin{figure}[ht]
  \begin{center}
    \includegraphics[trim=2.6in 4.9in 2.2in 0.8in,
      clip=true]{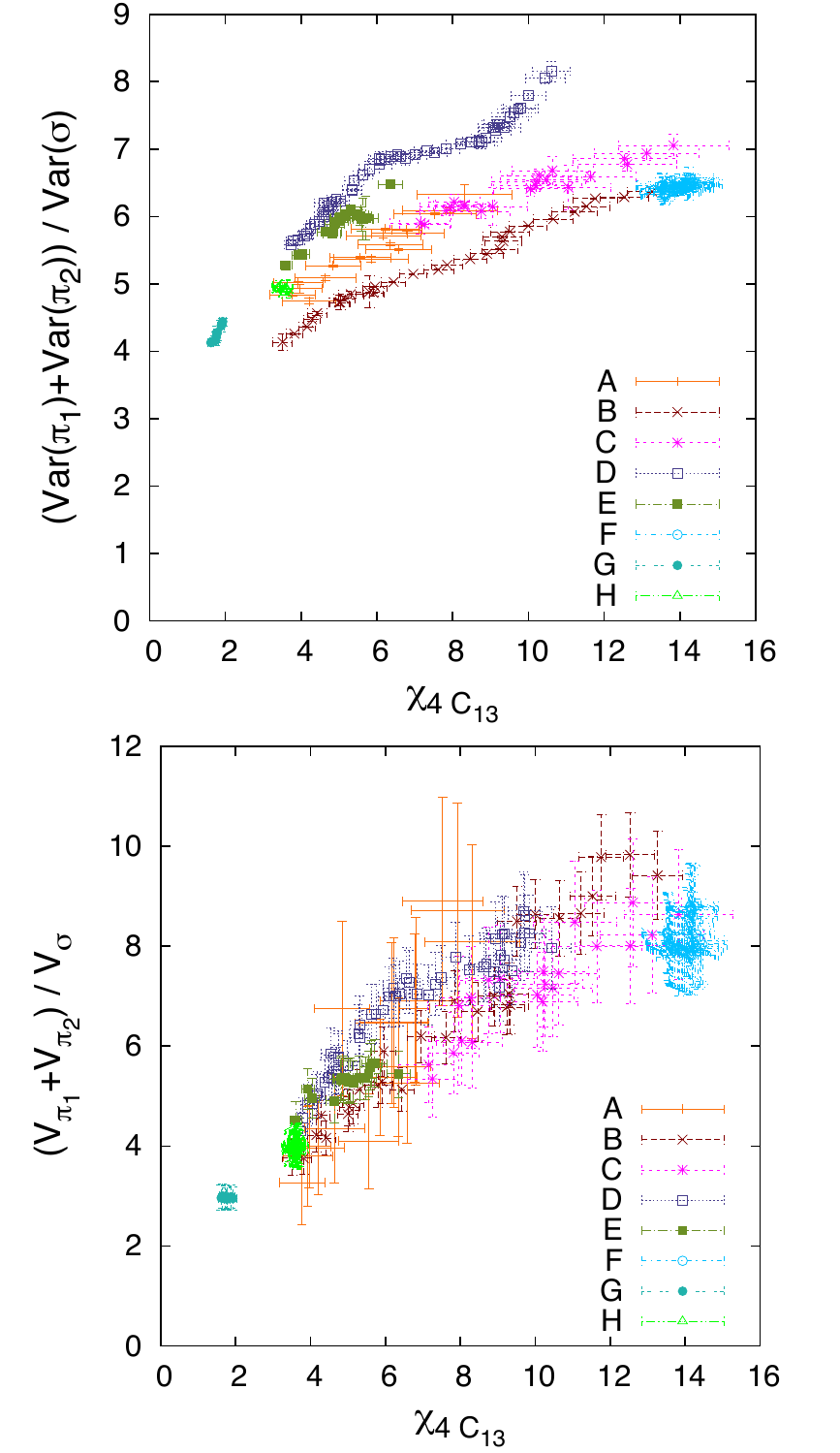}
  \end{center}
   \caption{(Color online) Anisotropy ratios between transverse and longitudinal
    fluctuations, plotted as functions of $\chi_{4, C(t_1,t_3)} \equiv
    \chi_4(t_1,t_3)$~\cite{Toninelli2005, Parsaeian2008}, which
    measures the strength of the dynamical heterogeneity. Plotted for
    all datasets and all times consistent with $C(t_1, t_3)=0.23$ and
    $C(t_1, t_2)= C(t_2, t_3)$. {\em Top panel:} Ratio between the
    variances of the transverse and longitudinal fluctuations. {\em
      Bottom panel:} Ratio between the correlation
    volumes~\cite{corr_vol} of the transverse and longitudinal
    fluctuations.}
  \label{fig:evolution}
\end{figure}

\begin{figure}[ht]
  \begin{center}
    \includegraphics[trim=0.9in 7.85in 2.8in 0.8in,
      clip=true]{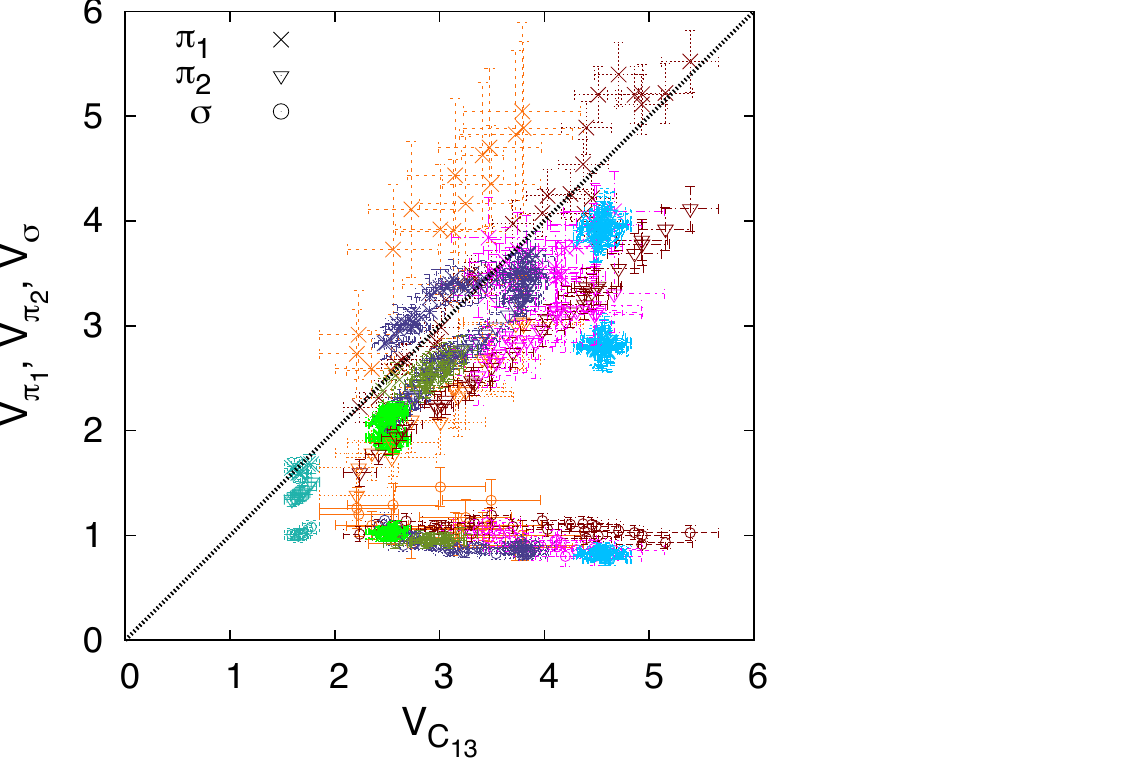}
  \end{center}
  \caption{(Color online) Normalized correlation volumes $V_{\pi_1}$ (crosses),
    $V_{\pi_2}$ (triangles), and $V_{\sigma}$ (circles) for the
    transverse fluctuations ${\pi}_{1 {\bf r}}, {\pi}_{2 {\bf r}}$ and
    the longitudinal fluctuations $\sigma_{\bf r}$, plotted as
    functions of the normalized correlation volume $V_{C_{13}} \equiv
    V_{C(t_1, t_3)}$
    of the dynamical heterogeneities~\cite{corr_vol}. Plotted
    for all datasets and all times consistent with $C(t_1, t_3)=0.23$
    and $C(t_1, t_2)= C(t_2, t_3)$. The color key for the datasets is
    as in Fig.~\ref{fig:evolution}.}
  \label{fig:corr-vol}
\end{figure} 

To test the hypothesis given by Eq.~(\ref{eq:C_goldstone}), we will
use the fact that for our data~\cite{Bouchaud1997}:
\begin{equation}
  C(t,t_w) \approx f \left( h(t)/h(t_w) \right),
  \label{eq:TR2}
\end{equation}
where $f$ can be fitted with a form such that $C(t,t_w)$ reduces to a
stretched exponential in the equilibrium case: $f(x)= q_{EA}
\exp\{-[\ln(x)/\theta_0]^{\beta}\}$~\cite{fit-parameters}. Here $q_{EA}$, $\beta$ and
$\theta_0$ are fitting parameters that vary little from one dataset to
another. However, 
the dependence of the $\alpha$ relaxation time $\tau$ on $t_w$ is
quite different in the different systems we
consider~\cite{triangular_long}, and this leads to different forms for
$h(t)$~\cite{Bouchaud1997}: for aging polymers
$h(t)=\exp[\ln^{\alpha}(t/t_0)]$, for aging particles
$h(t)=\exp[(t/t_0)^{\alpha}]$, and in equilibrium $h(t) =
\exp(t/t_0)$.
We define $\Phi_{ab} \equiv f^{-1}\left[C(t_a,t_b)\right]$, with $a, b
\in \{ 1, 2, 3 \}$. If Eq.~(\ref{eq:TR2}) is satisfied, we have
$\Phi_{ab} \approx h(t_a)/h(t_b)$, and we therefore obtain a {\em
  triangular relation\/}~\cite{Bouchaud1997} $ \Phi_{13} \approx
\Phi_{12} \Phi_{23}$. In terms of the
variables $X \equiv \Phi_{23}/\sqrt{\Phi_{13}}$ and $Y \equiv
\Phi_{12}/\sqrt{\Phi_{13}}$, this leads to the prediction that $1
\approx XY$, which is satisfied to a good approximation for all times
and all of our datasets~\cite{triangular_long}.

By using Eq.~(\ref{eq:TR2}), we now re-express our hypothesis,
Eq.~(\ref{eq:C_goldstone}), in the form $C_{\bf r}(t,t_w) = f \left(
{h_{\bf r}(t)}/{h_{\bf r}(t_w)} \right)$.  We now define $\Phi_{ab,
  {\bf r}} \equiv f^{-1}\left[C_{\bf r}(t_a,t_b)\right]$, with $a, b
\in \{ 1, 2, 3 \}$, $X_{\bf r}(t_1, t_2, t_3) \equiv \Phi_{23, {\bf
    r}}/\sqrt{\Phi_{13, {\bf r}}}$ and $Y_{\bf r}(t_1, t_2, t_3)
\equiv \Phi_{12, {\bf r}}/\sqrt{\Phi_{13, {\bf r}}}$, whose
fluctuations also encode the properties of the dynamical
heterogeneities. If the hypothesis in Eq.~(\ref{eq:C_goldstone}) is
satisfied, then $1 = X_{\bf r} Y_{\bf r}$, i.e., the relation holds
locally not just globally. Since time reparametrization symmetry is a
long time asymptotic effect associated with glassy behavior, we expect
that as the temperature becomes lower, the timescales become longer,
and the system becomes more glassy, the probability distribution
$\rho(X_{\bf r}, Y_{\bf r})$ should become anisotropic, and extend
mostly in the direction of the global curve $1 = X Y$ and not {\em
  away\/} from it. In other words, if we decompose the fluctuations
representing the dynamical heterogeneity into longitudinal and
transverse variables~\cite{Kardar2007}, the fluctuations of the
longitudinal variable ${\sigma_{\bf r}} \equiv \frac{1}{\sqrt{3}} \ln (
\Phi_{12, {\bf r}} \Phi_{23, {\bf r}} / \Phi_{13, {\bf r}} ) =
\frac{1}{\sqrt{3}} \ln (X_{\bf r} Y_{\bf r})$ should become weaker
than the fluctuations of the transverse variables ${\pi}_{1 {\bf r}} \equiv
\frac{1}{\sqrt{2}} \ln ( \Phi_{12, {\bf r}} / \Phi_{23, {\bf r}} ) =
\frac{1}{\sqrt{2}} \ln(Y_{\bf r}/ X_{\bf r})$ and ${\pi}_{2 {\bf r}} \equiv
\frac{1}{\sqrt{6}} \ln ( \Phi_{12, {\bf r}} \Phi_{23, {\bf r}}
\Phi_{13, {\bf r}}^2 )$.

In Fig.~(\ref{fig:all}) we show our results for $\rho(X_{\bf r},
Y_{\bf r})$. Because we are trying to detect collective fluctuations,
we coarse grain over moderately large regions, containing on average
125 particles. For each dataset, we find three triads of times $t_1 >
t_2 > t_3$ such that $(X(t_1, t_2, t_3), Y(t_1, t_2, t_3)) \approx
(0.8, 1.25), (1.00, 1.00),$ and $(1.25, 0.80)$ respectively. 
For each dataset and time triad, we show three contours of constant
probability density $\rho(X_{\bf r}, Y_{\bf r})$, respectively
enclosing $25\%$, $50\%$ and $75\%$ of the total probability. For
datasets A-D, with $T/T_c \sim 0.9$, the contours indeed follow the
curve $1 = XY$. This is more noticeable for the $25\%$ contour, which
encloses the most likely fluctuations, than for the $50\%$ and $75\%$
contours, which additionally include rarer events. For datasets E and
F, corresponding to $T/T_c = 1.1$, the contours are still anisotropic
and oriented along the direction of the global curve, but less so than
in A-D, while for G and H, corresponding to $T/T_c = 1.5$ the
fluctuations away from the global curve are the strongest. For the
higher temperatures, we find that the contours obtained in the aging
regime (F, H) are similar to the ones obtained in the equilibrium
regime (E, G) at the same temperatures~\cite{Parsaeian2009}.  These
results can be directly connected to the fact that, as the temperature
is increased, the separation of timescales is less pronounced, the
finite time corrections to the time reparametrization symmetry become
larger, and the effect of local time variable fluctuations become
weaker.

We now turn to a more quantitative analysis of the connection between
the transverse fluctuating variables $\pi_{1 {\bf r}}, \pi_{2 {\bf
    r}}$, the longitudinal fluctuating variables $\sigma_{\bf r}$, and
the dynamical heterogeneity. A more detailed version of this analysis
will be presented elsewhere~\cite{triangular_long}. Here we report
results for fixed $C(t_1,t_3) = 0.23$, but similar results are
obtained for other values of $C(t_1,t_3)$~\cite{triangular_long}. In
the top panel of Fig.~(\ref{fig:evolution}) we show the ratio between
the variances of the local transverse and longitudinal fluctuations as
a function of $\chi_{4, C(t_1,t_3)}$~\cite{Toninelli2005,
    Parsaeian2008, corr_vol}, which quantifies
  the strength of the dynamical heterogeneities. 
Similarly, in
the bottom panel of Fig.~(\ref{fig:evolution}) we plot the ratio
between the correlation volumes~\cite{corr_vol} of transverse and
longitudinal fluctuations as a function of $\chi_{4, C(t_1,t_3)}$. 
In both cases, we find that there is an anisotropy in favor of
the transverse fluctuations, which grows as the strength of the dynamical
heterogeneity increases. In particular, both ratios grow as the
temperature is decreased, and in the case of systems in the aging
regime, both ratios grow as the system relaxes, since 
$\chi_{4, C(t_1,t_3)}$ is a growing function of $t_w$ at fixed
$C(t_1,t_3)$~\cite{Parsaeian2008}. 

Our hypothesis is that the dynamical heterogeneity originates in the
Goldstone modes associated to fluctuations in the time
reparametrization, as described by Eq.~(\ref{eq:C_goldstone}).  We
thus expect that the correlation length of the dynamical heterogeneity
should be similar to the correlation lengths of the transverse
variables $\pi_1$ and $\pi_2$, and that the longitudinal variable
$\sigma$ should be short-range correlated. In
Fig.~(\ref{fig:corr-vol}), we show that this is indeed the case: the
normalized correlation volumes~\cite{corr_vol} $V_{\pi_1}, V_{\pi_2}$
for the transverse fluctuations are approximately proportional to
those for the dynamical heterogeneities, $V_{C_{13}}$, and in
particular they grow as the temperature is reduced or as aging systems
relax.  By contrast, the normalized correlation volume for longitudinal
fluctuations $V_{\sigma}$ is essentially constant for all systems,
temperatures and time regimes, and approximately equal to unity,
indicating that the spatial correlations of the variable $\sigma$ do not
extend beyond the coarse graining region. 


In conclusion, we have applied a stringent microscopic test for the
hypothesis that dynamical heterogeneity in structural glasses is
associated with the presence of {\em spatially correlated fluctuations
  in the time variables\/}, and we have found that all our results are
consistent with this hypothesis. We have used data from molecular
dynamics simulations of atomistic systems to apply the test, but the
same procedure can be applied to particle tracking data from
colloidal~\cite{Weeks2000} and granular systems~\cite{Keys2007}, and
slight modifications would allow the study of light
scattering~\cite{Cipelletti2003} or dielectric
noise~\cite{Russell2000} data. This opens the door to investigating
the possibility of a unified theoretical explanation of dynamical
heterogenity for molecular liquids, colloidal liquids and granular
systems. Our results highlight the advantages of studying dynamical
heterogeneity by probing fluctuations of {\em regions\/} of the
system, rather than probing {\em individual particle\/} fluctuations,
since the latter will necessarily contain both collective and
non-collective components that are difficult to separate cleanly. They
also highlight the fact that more complex correlations in the time
domain contain information that is useful for the understanding of
heterogeneous dynamical behavior.


H.~E.~C. thanks L.~Cugliandolo and C.~Chamon for suggestions and
discussions.  This work was supported in part by DOE under grant
DE-FG02-06ER46300, by NSF under grants PHY99-07949 and PHY05-51164,
and by Ohio University. Numerical simulations were carried out at the
Ohio Supercomputing Center.  H.~E.~C. acknowledges the hospitality of
the Aspen Center for Physics and the Kavli Institute for Theoretical
Physics, where parts of this work were performed.


\begin{thebibliography}{10}
\bibitem{corr_vol} We estimate normalized correlation volumes by the
  formula $V_{a} \equiv \frac{\chi_{4, a}}{V_{\rm coarse} {\rm Var}(a)}$,
  where $\chi_{4, a} \equiv V ( \langle A^2 \rangle - \langle A
  \rangle^2 )$, ${\rm Var}(a) \equiv \langle a^2 \rangle - \langle a
  \rangle^2$, $a$ is a local coarse grained variable, $A$ is the
  spatial average of $a$, $V$ is the total volume of the system, $
  V_{\rm coarse}$ is the coarse graining volume, and $\langle \cdots
  \rangle$ denotes an average over MD runs~\cite{triangular_long}.
\end{thebibliography}
\end{document}